\documentclass[aps,prl,twocolumn,superscriptaddress]{revtex4-1}

\usepackage{graphicx}
\usepackage{subfigure}
\usepackage{dcolumn}
\usepackage{bm}

\usepackage{epstopdf}
\usepackage{color}
\usepackage{hyperref}

\usepackage{colortbl}
\usepackage{multirow}
\usepackage{natbib}
\hypersetup{colorlinks=true, citecolor=blue, linkcolor=blue, urlcolor=blue}

\begin{document}


\title{Structure of self-assembled Mn atom chains on Si(001)}


\author{R. Villarreal}
\affiliation{Department of Quantum Matter Physics, University of Geneva, 24 Quai Ernest-Ansermet, CH-1211 Geneva 4, Switzerland}

\author{M. Longobardi}
\affiliation{Department of Quantum Matter Physics, University of Geneva, 24 Quai Ernest-Ansermet, CH-1211 Geneva 4, Switzerland}

\author{S. A. K\"{o}ster}
\affiliation{Department of Quantum Matter Physics, University of Geneva, 24 Quai Ernest-Ansermet, CH-1211 Geneva 4, Switzerland}

\author{Ch. J. Kirkham}
\affiliation{Division of Precision Science and Technology and Applied Physics, Graduate School of Engineering, Osaka University, 2-1, Yamada-oka, Suita, Osaka 565-0871, Japan}
\affiliation{London Centre for Nanotechnology and Department of Physics and Astronomy, University College London, London WC1E 6BT, United Kingdom}

\author{D. Bowler}
\affiliation{London Centre for Nanotechnology and Department of Physics and Astronomy, University College London, London WC1E 6BT, United Kingdom}

\author{Ch. Renner}
\altaffiliation{Corresponding author:\\Christoph.Renner@unige.ch}
\affiliation{Department of Quantum Matter Physics, University of Geneva, 24 Quai Ernest-Ansermet, CH-1211 Geneva 4, Switzerland}


\date{\today}

\begin{abstract}
Mn has been found to self-assemble into atomic chains running perpendicular to the surface dimer reconstruction on Si(001). They differ from other atomic chains by a striking asymmetric appearance in filled state scanning tunneling microscopy (STM) images. This has prompted complicated structural models involving up to three Mn atoms per chain unit. Combining STM, atomic force microscopy and density functional theory we find that a simple necklace-like chain of single Mn atoms reproduces all their prominent features, including their asymmetry not captured by current models. The upshot is a remarkably simpler structure for modelling the electronic and magnetic properties of Mn atom chains on Si(001).
\end{abstract}

\pacs{68.37.Ef, 68.37.Ps, 71.15.Mb, 73.20.-r}

\maketitle



Electrons confined to one dimension are expected to develop remarkable properties, both static and dynamic. Peierls transitions \cite{peierls55,haldane81}, collective spin and charge modes and Tomonaga Luttinger Liquid behaviour \cite{segovia99, bockrath99, blumenstein11} are among the predicted hallmarks of electrons in this extreme one dimensional (1D) quantum limit. Understanding 1D electrons also has technological implications owing to the drastic downscaling of devices and interconnects \cite{weber12}. However, the experimental realization of a truly 1D electronic system is a challenging endeavour. Individual atomic chains of limited length have been assembled atom by atom using scanning probes \cite{nilius02,hirjibehedin07,loth12}. Much longer single atom chains can be synthesized in significant numbers by self-assembly along step edges \cite{segovia99, gambardella02, erwin10} and on flat terraces \cite{oncel05, blumenstein11}. 

Modelling and understanding the electronic properties of atomic chains requires a detailed knowledge of their structure. While this is self-evident for chains constructed using a scanning probe, it has in some cases proven much more challenging to determine the structure of self-assembled atomic chains \cite{owen02,vanhouselt08}. Consequently, the interpretation of their spectroscopic signatures is often difficult and controversial. Progress in modelling their electronic properties is directly linked to advances in their structural analysis, especially the positive identification and discrimination of structural and electronic features in the scanning tunneling microscope (STM) images \cite{yeom05, heimbuch12, aulbach13, park14}.

\begin{figure}
\includegraphics[width=\columnwidth]{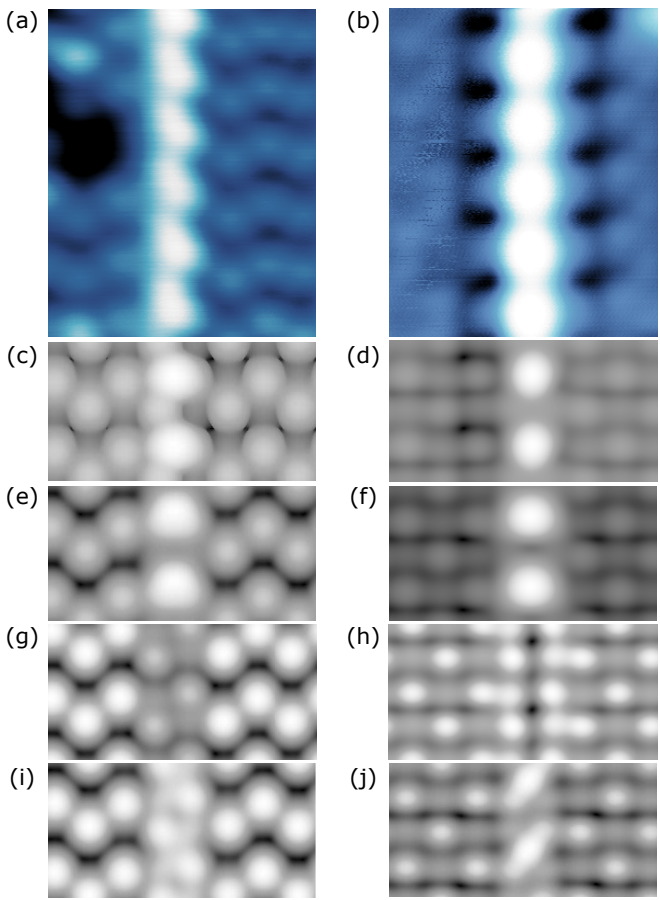}
\caption{\label{fig:stm}High resolution 22 $\times$ 24 \AA$^2$  STM micrographs (a-b) and selected DFT simulations: H' model (c-d), Wang model \cite{wang10} (e-f), H model (g-h) and C model (i-j). Filled states images are shown on the left (STM: -2.5 V, 150 pA; DFT simulations at -1 V). Empty states images are shown on the right (STM: 1.8 V, 150 pA; DFT simulation at 1 V). See text and Fig.~\hyperref[fig:afmmodel]{2(b)} for details about the DFT models.}
\end{figure}

Several metal atoms have been found to form dimers and dimer chains oriented perpendicular to the silicon dimer rows on the Si(001) surface at low coverage \cite{nogami91}. Depending on coverage and growth temperature, other structures are observed, including island growth and alloying. Mn was recently found to form atomic chains on Si(001) \cite{liu08}. While they also run perpendicular to the Si dimer rows, their precise structure remains unsolved. Several models, from single-atom to trimer systems, have been proposed \cite{wang10,sena11,niu12}. But they all fail to reproduce a striking asymmetry of the Mn chains observed in filled states STM images (Fig.~\hyperref[fig:stm]{1(a)}) \cite{liu08,nolph11,fuhrer12}. Here, we combine STM, non-contact atomic force microscopy (NC-AFM) and density functional theory (DFT) modelling to unveil their microscopic structure.  


Manganese atom chains studied here were self-assembled on clean p-type (boron doped, 0.1~$\Omega$cm) and n-type (arsenic doped, 0.01~$\Omega$cm) reconstructed Si(001) surfaces. The deposition was performed in UHV (base pressure $\sim$10$^{-11}$ mbar) at room temperature prior to the STM and AFM investigations. All scanning probe images reported here were obtained in UHV at 78 K using an Omicron LT-STM. Electrochemically etched W tips and cut Pt/Ir tips were used for the constant current STM measurements. The NC-AFM images were acquired at constant frequency shift ($\Delta f$) using commercial qPlus tuning-fork sensors \cite{giessibl00(2)} that were grounded to avoid artefacts \cite{wutscher12}. 

The STM images and adsorption energies were modeled by DFT using the Vienna \emph{Ab initio} Simulation Package (VASP) version 4.6.34 \cite{kresse96}. In all cases the core electrons were described by the projector augmented wave (PAW) method \cite{kresse99}. The Si(001) surface was represented by a periodically repeated eight layer slab, with a p(2$\times$2) reconstructed surface layer consisting of two rows of 8 Si dimers. The bottom Si layer was terminated by H atoms in a dihydride structure. Prior to the full structure relaxations the bottom two Si layers and the H atoms were allowed to relax whilst keeping all other Si atoms fixed, to optimize the Si-H bond lengths. For all subsequent calculations both the H atoms and the bottom two layers of Si atoms were fixed, in order to simulate a bulk like environment. All calculations were spin polarized. The PW91 exchange-correlation functional \cite{perdew92} was used, with a plane wave cut off of 320 eV and the valence state 3p$^6$3d$^5$4s$^2$ for Mn. For these calculations a (2$\times$1$\times$1) Monkhorst-Pack k-mesh was used, with the largest number of k-points along the direction of the Mn chain. $E_{ads}/Mn = (E_{surface + Mn} - E_{surface} - nE_{isolated\ Mn})/n$ (where n is the number of Mn atoms) was used to calculate the adsorption energies of Mn atoms on the surface. DFT+U calculations \cite{anisimov91} were performed using Coulomb and exchange parameters of U = 4.2 eV and J = 1 eV (in accordance with prior calculations \cite{sena11}). STM images were simulated using the Tersoff-Hamman method, as implemented in bSKAN33 \cite{hofer03}. 

Two high resolution occupied and empty states STM micrographs of a Mn chain on Si(001) are shown in Figs.~\hyperref[fig:stm]{1(a)} and \hyperref[fig:stm]{1(b)}, respectively. The p(2$\times$2) buckled dimer reconstruction of the silicon background running horizontally on both sides of the vertical Mn chain is well resolved. In empty states images (Fig.~\hyperref[fig:stm]{1(b)}), the Mn chain appears as a regular pearl-necklace like atomic assembly, very similar to other atomic chains on Si(001). The dark atomic rows along both sides reflect a different Si dimer configuration discussed later. Filled states images of the Mn chain (Fig.~\hyperref[fig:stm]{1(a)}) present a very different profile with a singular asymmetric shape: one edge is straight and the opposite one is crenellated. This unusual appearance has prompted the development of different theoretical chain models involving up to three atoms per unit cell, but none has been able to capture the negative sample bias asymmetry.  

\begin{figure}
\includegraphics[width=\columnwidth]{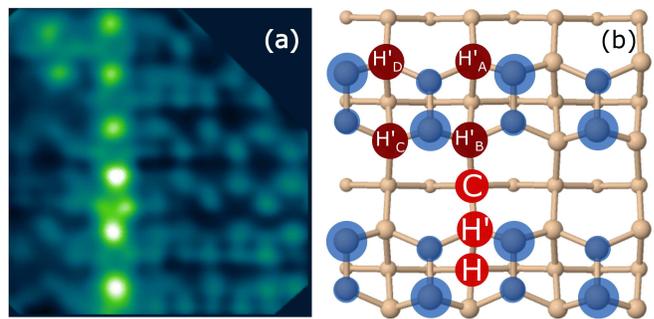}
\caption{\label{fig:afmmodel} (a) 33 $\times$ 27 \AA$^2$ NC-AFM image of a Mn chain at constant $\Delta f$ = -17 Hz and oscillation amplitude ($A$) = 1 nm. The bright vertical atom row is the Mn chain running perpendicular to the Si(001) dimer rows in the background. (b) Ball and stick model of adsorption sites investigated for Mn atoms on Si(001) p(2$\times$2) surface as introduced in Ref.~\cite{sena11}. Small and large blue dots represent down and up Si dimer atoms, respectively. Red dots show the Mn adsorption sites considered.}
\end{figure}

The contrast of STM images is a complex convolution of actual topography and local electron density of states. Periodic atomic scale features may thus not necessarily correspond to atomic lattice sites. Atomic force microscopy provides a different contrast and has been demonstrated to be complementary to STM in resolving surface structures \cite{eguchi04,setvin12,emmrich15}. A high resolution NC-AFM image of a Mn chain at 78~K is shown in Fig.~\hyperref[fig:afmmodel]{2(a)}. The bright vertical structure running perpendicular to the horizontal Si dimer rows is a Mn chain. This NC-AFM image clearly suggests a chain structure consisting of regularly spaced single atoms rather than a more complex dimer or trimer structure. It further gives direct clues to the position of each Mn atom: it is sitting close to one edge of the Si dimer row and half-way between two adjacent Si dimers. This observation favours the H' lattice site for the Mn atom (Fig.~\hyperref[fig:afmmodel]{2(b)}), excluding several other sites, in particular the C and H sites considered in Refs.~\cite{wang10,sena11}. As shown in Figs.~\hyperref[fig:stm]{1(c-d)}, the H' chain indeed reproduces best the Mn chain contrast seen by STM. It additionally suggests that its asymmetric appearance is induced by the Si buckling. The Wang trimer \cite{wang10}, which consists of H'-H-H' motifs, reproduces the empty states image well, but fails to show the characteristic filled states asymmetry as it alters the buckling pattern of the underlying Si dimers (Fig.~\hyperref[fig:stm]{1(e)}). We found the trimer contrast to be very sensitive to different settings in the calculations, with the relative heights of the Mn atoms next to the trench varying significantly. But none of the settings reproduced the striking filled state asymmetric appearance.

\begin{figure}
\includegraphics[width=\columnwidth]{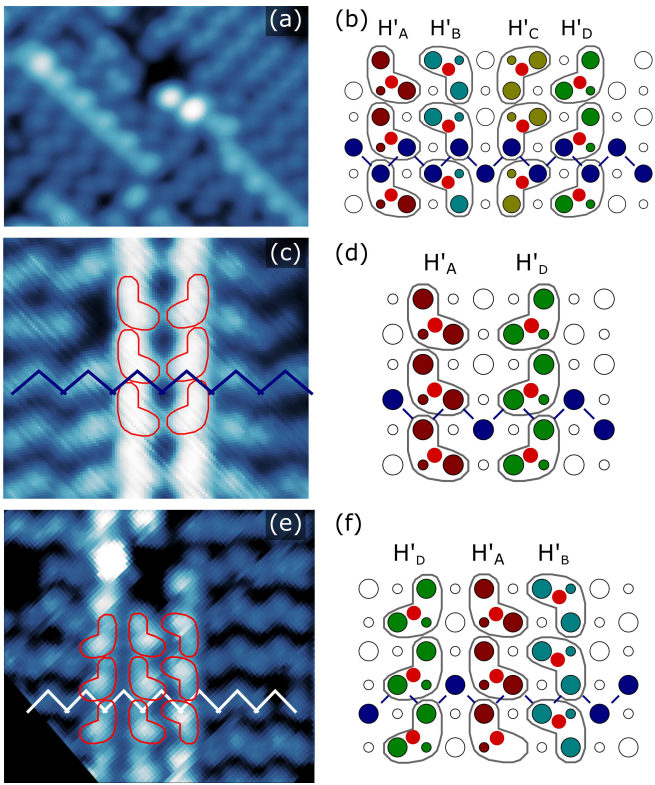}
\caption{\label{fig:asymconf} Filled states constant current STM images of Mn chains on Si(001) with different configuration and cartoon models. (a) 34 $\times$24 \AA$^2$ image at -3 V and 200 pA. (b) Model representation of the four possible H' chain configurations. The units building the crenellated geometry are outlined in gray with Mn atoms in red. (c-f) Show three observed closest configurations of H' chains and their corresponding model descriptions. The zig-zag highlights the registry of the H' chains on the Si dimer rows. (c) 37 $\times$ 30 \AA$^2$ image of  H'$_A$ and H'$_D$ chains at -2.5 V and 80 pA. (d) 60 $\times$ 43 \AA$^2$ image of H'$_D$, H'$_A$ and H'$_B$ at -3.3 V and 100 pA.}
\end{figure}

We observe four distinct orientations of the Mn chain asymmetry in our low temperature STM micrographs (Fig.~\hyperref[fig:asymconf]{3}). They correspond to the four ``\textit{wire types}" identified at room temperature by Fuhrer \textit{et al.} \cite{fuhrer12}. The H' model provides a straightforward explanation for these observations. Indeed, there are four non-equivalent H' sites on the p(2$\times$2) reconstructed silicon surface corresponding to each of the four Mn chain orientations, as illustrated in Fig.~\hyperref[fig:asymconf]{2(b)}. In addition, the H' model accounts for the closest Mn chain separations observed by STM depending on their relative orientations (Figs.~\hyperref[fig:asymconf]{3(c-f)}). Fuhrer \textit{et al.} \cite{fuhrer12} report flipping of the chain asymmetry during STM scanning which they relate to changes in the Mn trimer buckling or to changes in the Si buckling or a combination thereof. The H' model provides an elegant explanation for the chain flipping associated with Mn trimer flipping in terms of Mn atoms diffusing perpendicular to the Si dimer rows between configurations H'$_A$ and H'$_B$ or H'$_C$ and H'$_D$ (Fig.~\hyperref[fig:asymconf]{3(b)}). 

\begin{figure}
\includegraphics[width=\columnwidth]{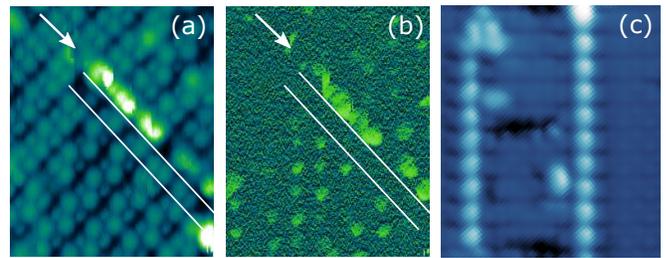}
\caption{\label{fig:strainspm} (a) 36 $\times$ 50 \AA$^2$ topographic NC-AFM image of a Mn chain measured at constant $\Delta f$ = -9.7 Hz and $A$ = 1 nm.  (b) Dissipation signal corresponding to image (a). Increased dissipation is measured over Si down atoms and Mn atoms. (c) 54 $\times$ 61 \AA$^2$ STM constant current image of Mn chains on Si(001) surface at 1.5 V and 200 pA. The white arrows point at the Mn chain and the white lines delimit static dimers along the Mn chain.}
\end{figure}

We do not observe the flipping between Mn chain configurations reported at room temperature by Fuhrer \textit{et al.} \cite{fuhrer12} in our STM data. This is likely a direct consequence of the lower temperature (78 K) of our experiments. On the other hand, low temperature NC-AFM imaging does provide insight into the Mn diffusion potential associated with this flipping. Depending on tip condition and scanning parameters, single Mn atoms appear either well resolved (Fig.~\hyperref[fig:afmmodel]{2(a)}) or blurred (Fig.~\hyperref[fig:strainspm]{4(a)}). The blurring is independent of scanning direction and is the result of Mn atoms moving along a specific direction linking opposite H' sites. The NC-AFM image convincingly shows that the motion is confined within a given Si dimer row. There is no hopping observed along or between Si dimer rows. The emerging picture of Mn atoms hopping from one edge of a given Si dimer row to the opposite one is perfectly compatible with the flipping observed by Fuhrer \textit{et al.} \cite{fuhrer12} between configurations H'$_A$ and H'$_B$ or H'$_C$ and H'$_D$. In our low temperature experiment, the energy for the hopping is provided by the AFM tip, as directly seen in the increased dissipation on the Mn chain (Fig.~\hyperref[fig:strainspm]{4(b)}). This image again shows the Mn hopping to be confined within a Si dimer row, excluding interdimer hopping. It moreover excludes the diffusion of Mn chain atoms along the Si dimer rows. Hence, flipping between configurations H'$_A$ and H'$_D$ or H'$_B$ and H'$_C$ are only possible in combination with a flipping of the neighbouring Si buckling, as reported by Fuhrer \textit{et al.} \cite{fuhrer12}.

The Si dimers closest to the Mn chain look different from the rest of the surface in empty states STM micrographs (Figs.~\hyperref[fig:stm]{1(b)} and \hyperref[fig:stm]{4(c)}). Similar features along Bi nanolines on Si(001) have been explained in terms of local strain \cite{owen03}. The NC-AFM dissipation signal in Fig.~\hyperref[fig:stm]{4(b)}) is enhanced over the Mn and down Si atoms. The latter has been assigned to reversible buckled dimer flipping induced by the AFM tip \cite{hoffmann01,bamidele12}. The absence of dissipation at the Si atoms along the Mn chains is compatible with a locally stiffer Si dimer configuration. This may explain the absence of Mn diffusion perpendicular to the Mn chain and may contribute the Mn chain growth.

\begin{table}[b]
\caption{Adsorption energies for Mn single atoms and chains on the Si(001) reconstructed surface. The sites are depicted in Fig.~\hyperref[fig:afmmodel]{2(b)} and the Wang structure is the one used in Ref.~\cite{fuhrer12}. *Energy corresponding to single Wang trimer.}
 \label{table:1}
\begin{ruledtabular}
\begin{tabular}{|c|cccc|}
Structure & Single-atom & Chain & Single-atom +U & Chain +U\\
\hline
 H & -2.75 & -2.68 & -1.69 & -1.69\\
 H' & -2.51 & -2.51 & -1.86 & -1.85\\
 C & -1.59 & -1.75 & - & -1.11\\
Wang & -2.64* & -3.06 & - & -1.79\\
2H' & - & -2.88 & - & -2.11
\end{tabular}
\end{ruledtabular}
\end{table}

H', H and C are three possible Mn adsorption sites compatible with the single protrusion in our scanning probe images. However, only H' is located on the edge of the Si dimer row in agreement with the NC-AFM image shown in Fig.~\hyperref[fig:afmmodel]{2(a)}. H and C are centred on and between the Si dimer rows, respectively, which is not compatible with our data. Simple DFT calculations (Table~\hyperref[table:1]{1}) find that the H' site is one of the most stable Mn adsorption sites, though slightly less stable than the H site.  Addition of a local on-site shift, using DFT+U \cite{sena11}, changes the ordering and makes the H' site most stable, while the simulated STM is hardly altered. However, our calculations show no energetic gain in forming a chain of atoms located at the H' site on one side of the dimer row, either with or without Hubbard U. The relative stabilities of different structures depend on whether or not U is included: with U, a chain of pairs of atoms in both H' sites is most stable, while single atoms in one H' site are next most stable, better than the Wang trimer model. The H' dimer is especially notable given the high mobility of Mn across the dimer rows (with a mere energy barrier of $\sim$0.16 eV per Mn), in agreement with our experimental observations (Figs.~\hyperref[fig:stm]{4(a-b)}). Nonetheless, Mn dimers have not been experimentally observed on the surface and all these structures produce a poor match to experimental STM when compared to the simple H' model. There must be some ingredients critical to the formation of H' chains which are not yet identified and not included in our model, such as sub-surface or in-surface atoms (which could be hidden from STM and NC-AFM), local charging or strain effects.

In summary, combining scanning tunneling microscopy, non-contact atomic force microscopy and density functional theory modelling, we have been able to solve the atomic structure of Mn chains self-assembled on the Si(001) surface. We are confident that the STM contrast in the Mn chain is produced by Mn atoms in the H' position. However, there are some subtleties in the chain structure that are not yet clear, in particular relating to why the chains are straight, and why H' dimers do not form. In all structures we have modelled, we have found either good energetics with poor match to STM, or less stable structures with good match to STM. Nevertheless, we find that a simple pearl-necklace like assembly of Mn atoms adsorbed on the H' site (Fig.~\hyperref[fig:afmmodel]{2(b)}) reproduces best all the Mn chain features reported so far, including their asymmetric appearance in occupied states STM micrographs. The structure we propose is much simpler than the currently prevailing Wang trimer model involving three Mn atoms per chain unit \cite{wang10}. The unusual asymmetric outline which has motivated the more complex trimer model is a consequence of the buckled Si dimer background. Our refined structural model provides a new basis for modelling the electronic and magnetic properties of these chains.

We acknowledge discussions with James H. G. Owen, A. Scarfato and H. Zandvliet. We thank G. Manfrini for his technical support. This work was financially supported by the Swiss National Science Foundation. D. R. Bowler and Ch. Kirkham were partly funded by the EPSRC under the COMPASSS programme grant number EP/H026622/1 and acknowledges computational facilities at the London Centre for Nanotechnology.


\end{document}